\documentclass[
   pre,aps,
   amsfonts,
   amssymb,
   amsmath,
   superscriptaddress,
   eqsecnum,
   showpacs,
   preprintnumbers,
   byrevtex]{revtex4}
\usepackage{color}
\usepackage{graphicx}

\begin{document}

\newcommand{\vphi}{\varphi}
\newcommand{\bq}{\begin{equation}}
\newcommand{\be}{\begin{equation}}
\newcommand{\ba}{\begin{eqnarray}}
\newcommand{\eq}{\end{equation}}
\newcommand{\ee}{\end{equation}}
\newcommand{\ea}{\end{eqnarray}}
\newcommand{\tchi} {{\tilde \chi}}
\newcommand{\tA} {{\tilde A}}
\newcommand{\sech} { {\rm sech}}
\newcommand{\pstar}{\mbox{$\psi^{\ast}$}}
\newcommand {\bPsi}{{\bar \Psi}}
\newcommand {\bpsi}{{\bar \psi}}
\newcommand {\bphi}{{\bar \phi}}
\newcommand {\tu} {{\tilde u}}
\newcommand {\tv} {{\tilde v}}
\newcommand{\dq}{{\dot q}}
\preprint{LA-UR 16-21471} 

\title{Approximate Analytic Solutions to Coupled Nonlinear Dirac Equations}
\author{Avinash Khare}
\email{khare@physics.unipune.ac.in}
\affiliation{Physics Department, Savitribai Phule Pune University, Pune 411007, India}

\author{Fred Cooper}
\email{cooper@santafe.edu}
\affiliation{Santa Fe Institute, Santa Fe, NM 87501, USA}
\affiliation{Theoretical Division and Center for Nonlinear Studies, 
Los Alamos National Laboratory, Los Alamos, New Mexico 87545, USA}
\author{Avadh Saxena}
\email{avadh@lanl.gov}
\affiliation{Theoretical Division and Center for Nonlinear Studies, 
Los Alamos National Laboratory, Los Alamos, New Mexico 87545, USA}

\begin{abstract}
We consider the coupled nonlinear Dirac equations (NLDE's)  in 1+1 dimensions with 
scalar-scalar  self interactions 
$\frac{ g_1^2}{2} ( {\bpsi} \psi)^2 + \frac{ g_2^2}{2} ( {\bphi} \phi)^2 + g_3^2  ({\bpsi} \psi) ( {\bphi} \phi)$ as well as vector-vector interactions of the form
$\frac{g_1^2 }{2} (\bpsi \gamma_{\mu}  \psi)(\bpsi \gamma^{\mu} \psi)+ \frac{g_2^2 }{2} 
(\bphi \gamma_{\mu} \phi)(\bphi \gamma^{\mu} \phi) + g_3^2 (\bpsi \gamma_{\mu} \psi)(\bphi \gamma^{\mu}  \phi ). $
Writing  the two components of the assumed  solitary wave solution of these equation in the form   $\psi = e^{-i \omega_1 t} \{R_1 \cos \theta,   R_1 \sin \theta \}$, $\phi = e^{-i \omega_2 t} \{R_2 \cos \eta,   R_2\sin \eta \}$, and assuming 
that $ \theta(x),\eta(x)$ have the {\it same}  functional form they had 
when $g_3$=0, which is an approximation consistent with the conservation 
laws, we then find approximate analytic solutions for $R_i(x)$  which are 
valid  for small values of $g_3^2/ g_2^2  $ and  $g_3^2/ g_1^2$.
In the nonrelativistic limit we show that both of these coupled models 
go over to the same coupled nonlinear Schr\"odinger equation for which 
we obtain two exact pulse solutions vanishing at $x \rightarrow \pm \infty$.

\end{abstract}


\maketitle


\section{Introduction}

The nonlinear Dirac (NLD) equation  in $1+1$ dimensions \cite{iva} 
has a long history and 
has emerged as a useful model in many physical systems such as extended 
particles \cite{fin,ffk,hei}, the gap solitons in nonlinear optics 
\cite{bar}, light solitons 
in waveguide arrays and experimental realization of an optical analog for 
relativistic quantum mechanics \cite{lon,dre,tra}, Bose-Einstein condensates in 
honeycomb optical lattices \cite{had}, phenomenological models of quantum 
chromodynamics \cite{fil}, as well as matter influencing the evolution of the 
universe in cosmology \cite{sah}. Further, the multi-component BEC order 
parameter has an exact spinor structure and serves as  the bosonic 
analog to the relativistic electrons in graphene. 
To maintain the Lorentz invariance of the NLD 
equation, the self interaction Lagrangian is built using the bilinear 
covariants. Of special interest are scalar bilinear covariant and vector bilinear 
covariant which have particularly attracted a lot of attention. 

Classical solutions of nonlinear field equations have a long history as a 
model of extended particles~\cite{sol}. In 1970, 
Soler proposed that the self-interacting 4-Fermi theory was an 
interesting model for extended fermions.  Later, Strauss and Vasquez 
\cite{str}  were able to study the stability of this model under 
dilatation and found the domain of stability for the Soler solutions. 
Solitary waves in the 1+1 dimensional nonlinear Dirac equation have 
been studied \cite{lee,nog} in the past in case 
the nonlinearity parameter $k=1$,  i.e. the massive Gross-Neveu~\cite{gro} (with
$N=1$, i.e. just one localized fermion) and the massive Thirring~\cite{thi} 
models. In those studies it was found that these equations have solitary wave 
solutions for both scalar-scalar (S-S) and vector-vector (V-V) interactions. 
The interaction between solitary waves of different initial charge was 
studied in detail  for the S-S case 
in the work of Alvarez and Carreras \cite{alv} 
by Lorentz boosting the static solutions and allowing them to scatter. 

In a previous paper \cite{coo} we extended the work of these preceding  
authors to the case where the nonlinearity was taken to an arbitrary 
power $\kappa$ for both the scalar-scalar and vector vector couplings and 
were able to find solitary wave solutions for an 
arbitrary nonlinearity parameter $\kappa$. In this paper we will extend the 
previous models in a new direction by looking for solitary wave solutions to 
the problem of two coupled NLDE's  and considering the scalar-scalar coupling as 
well as the vector-vector coupling between the two fields.  Our strategy is to 
write the components of the two Dirac equations for solitary waves as  
$\{R _i(x) \cos \theta_i(x), R_i (x) \sin \theta_i(x) \} e^{-i \omega_i t}$ 
and then assume that the conservation law for linear momentum is satisfied 
independently for $i=1,2$. This assumption is equivalent to saying that  
$ \theta(x),\eta(x)$ have the {\it same}  functional form they had when 
$g_3$=0. Once one makes that assumption we obtain an analytic expression for 
$R_i(x)$  which we then show approximately solves the differential equation 
for $R_i(x)$.  The one situation which restricts the validity of this solution 
occurs in the scalar-scalar interaction case when one of the solitary wave 
solutions (when $g_3=0$) is of a double humped variety.  In that case the 
solution is valid only when the dimensionless coupling constants 
$g_3^2/ g_2^2  $ and  $g_3^2/ g_1^2$ are $\leq 1/100$. Otherwise the 
approximate analytic solutions we have found seem to be numerically accurate 
in both the scalar-scalar as well as the vector-vector coupled NLD equation 
as long as the two dimensionless constants are $\leq 1/10$.

\section{scalar-scalar interactions}
We are interested in  solitary wave solutions of the coupled nonlinear Dirac equations (NLDEs) given by 
\bq\label{1}
 (i \gamma^{\mu} \partial_{\mu} - m_1) \psi +g_1^2 (\bpsi  \psi)  \psi 
+ g_3^2 (\bphi  \phi)  \psi = 0\,,
\eq
\bq\label{2}
(i \gamma^{\mu} \partial_{\mu} - m_2) \phi +g_2^2 (\bphi  \phi)  \phi 
+ g_3^2 (\bpsi  \psi)  \phi = 0\,. 
\eq
We can eliminate one of the coupling constants by rescaling the fields, that 
is if we let $\psi \rightarrow  \psi/g_1$, $\phi \rightarrow  \phi/g_2$,  
so that  there are two independent dimensionless coupling  constants  
\bq\label{3}
g_{32}^2= g_3^2/ g_2^2  ,~~ g_{31}^2=g_3^2/ g_1^2.
\eq
as we will discover later. The field  equations can be derived from the 
Lagrangian
\ba\label{4}
&&\mathcal{L}  =  \bpsi (i \gamma^{\mu} \partial_{\mu} - m_1) \psi + \frac{g_1^2 }{2} (\bpsi  \psi) ^2    \nonumber \\
&& + \bphi (i \gamma^{\mu} \partial_{\mu} - m_2) \phi +\frac{g_2^2 }{2} (\bphi  \phi) ^2 + g_3^2 (\bpsi  \psi)(\bphi  \phi ) \nonumber \\
&&= \bpsi (i \gamma^{\mu} \partial_{\mu}- m_1) \psi + \bphi (i \gamma^{\mu} \partial_{\mu} - m_2) \phi + \mathcal{L} _{int} .
\ea
We notice the Lagrangian is symmetric under the interchange $\psi \rightarrow  \phi, m_1 \rightarrow m_2$ and $g_1 \rightarrow g_2$.

We next choose the following representation of the $\gamma$ matrices:
\bq\label{5}
 \gamma^0 =\sigma_3, ~~~ i \gamma_1 = \sigma_2 , 
 \eq
 where the $\sigma_i$ are the usual Pauli spin matrices.

In the rest frame we assume that  the two components of the  solutions  can 
be written as 
 \ba\label{6}
\psi(x) &&  =  \left(  \begin{array} {cc}
      A(x) \\
       ~B(x) \\ 
   \end{array} \right) e^{-i \omega_1 t} =
R_1(x) \left(\begin{array}{c}\cos \theta(x) \\  \sin \theta(x) \end{array}\right)  e^{-i \omega_1 t} , \nonumber \\
\phi(x) &&  =  \left(  \begin{array} {cc}
      C(x) \\
       ~D(x) \\ 
   \end{array} \right)  e^{-i \omega_2 t}  =
R_2(x) \left(\begin{array}{c}\cos \eta(x) \\ \sin \eta(x) \end{array}\right)  e^{-i \omega_2 t} . 
\ea
In the absence of interactions ($g_3=0$), the solutions are of two types 
\cite{coo}.  When $1 > \omega/m > \omega_c /m$ then the solutions are 
single humped as 
they are always in the case of  vector-vector  interactions discussed below.  
However  for the case $1 > \omega_c/m > \omega /m$ the solutions are double  
humped and in that regime if the solutions when $g_3=0$ are of two different 
types, then we will find the approximate solutions we obtain are only 
valid for very small $g_{3i}^2 \leq 1/100$.
In component form these two coupled NLDEs can be written as 
\ba\label{7}
&&\partial_x A + (m_1 +\omega_1) B - g_1^2 (A^2-B^2)  B - g_3^2 (C^2-D^2)  B  =0, \nonumber \\ 
&&\partial_x B + (m_1-\omega_1) A -  g_1^2 (A^2-B^2)  A - g_3^2 (C^2-D^2)  A =0, \nonumber \\
&&\partial_x C + (m_2 +\omega_2) D - g_2^2 (C^2-D^2)  D - g_3^2 (A^2-B^2)  D  =0, \nonumber \\ 
&&\partial_x D + (m_2 +\omega_2) C - g_2^2 (C^2-D^2)  C - g_3^2 (A^2-B^2)  C  =0 . 
\ea
These are symmetric under the interchange $ \{ A, B\} \rightarrow  \{ C, D\} ,m_1 \rightarrow m_2$ and $g_1 \rightarrow g_2$.
These four  equations  can also be written if we let $y_i=R_i^2(x)$ as:
\ba\label{8} 
\frac{dy_1}{dx} && = 2 [g_1^2 y_1^2  ( \cos 2 \theta) + g_3^2 y_1 y_2   
( \cos 2 \eta) -  y_1 m_1] \sin 2 \theta , \nonumber \\
\frac{dy_2}{dx} && = 2 [g_2^2 y_2^2  ( \cos 2 \eta) + 2 g_3^2 y_1 y_2   
( \cos 2 \theta) - y_2 m_2] \sin 2 \eta  ,
\ea
and
\ba  \label{9}
\frac{d \theta}{dx} &&=  g_1^2 y_1  \cos^2   2 \theta +  g_3^2 y_2  \cos   2 \theta  \cos   2 \eta- m_1 \cos 2 \theta + \omega_1 , \nonumber  \\
\frac{d \eta}{dx} &&=  g_2^2 y_2 \cos^2   2 \eta +  g_3^2 y_1  \cos   2 \theta  \cos   2 \eta- m_2 \cos 2 \eta + \omega_2 .  
\ea
We can rewrite these equations in terms of the two dimensionless coupling 
constants by scaling $y_1 \rightarrow y_1/g_1^2$, $y_2 \rightarrow y_2/g_2^2$.

%

\subsection{Conservation Laws}

We have that energy and momentum are conserved, namely
\bq\label{10}
\partial _\mu T^{\mu \nu} =0\, , 
\eq
where the energy-momentum tensor is defined as 
\bq\label{11}
T_{\mu \nu} = i \bpsi \gamma_\mu \partial_\nu \psi 
+ i \bphi  \gamma_\mu \partial_\nu \phi - g_{\mu \nu}  \mathcal{L}\,.
\eq
and $\mathcal{L}$ is given by Eq. (\ref{4}).
From total momentum conservation,  we find, just like for the single field NLDE, that for a solution that vanishes at $\pm \infty$  we have 
\bq\label{12}
T_{10}= \omega_1 \bpsi \gamma_1 \psi + \omega_2 \bphi \gamma_1 \phi =0\, 
\eq
and also
\bq \label{13}
T_{11} = \omega_1 \psi^\dag \psi - m_1 \bpsi \psi + \omega_2 \phi^\dag \phi 
- m_2 \bphi \phi + \mathcal{L}_{int}=0\,.
\eq
Multiplying Eq. (\ref{1}) on the left by $\bpsi$ and Eq. (\ref{2}) on the 
left by $\bphi$  and adding those two equations and then using  Eq. (\ref{13}) 
to eliminate the interaction terms of $\mathcal{L}_{int}$, we then obtain 
the equation:
\bq\label{14}
\omega_1 \psi^\dag \psi - m_1 \bpsi \psi + i \bpsi \gamma_1 \partial_1 \psi 
+\omega_2 \phi^\dag \phi -m_2 \bphi \phi + i\bphi \gamma_1 \partial_1 \phi =0\,,
\eq
which becomes using our ansatz
\bq\label{15}
R_1^2 \left( \frac {d \theta}{dx} + \omega_1 - m_1 \cos 2 \theta \right) 
+ R_2^2 \left( \frac{ d \eta}{dx} + \omega_2 - m_2 \cos 2 \eta \right) = 0\,.
\eq

One also has that energy is conserved.  The energy density is given by 
\bq\label{16}
T_{00}(x) = m_1 \bpsi \psi + m_2 \bphi \phi = m_1R_1^2  \cos 2 \theta 
+ m_2 R_2^2 \cos 2 \eta\,,
\eq
with the total energy being conserved
\bq\label{17}
E = \int_{- \infty}^{\infty} dx  \left(m_1 R_1^2  \cos 2 \theta 
+ m_2  R_2^2\cos 2 \eta \right)\,.
\eq

The other conserved quantities are charges $Q_{\psi}$ and $Q_{\phi}$ 
defined by
\be\label{18}
Q_{\psi} = \int_{-\infty}^{+\infty} dx~\psi^{\dag} \psi\,,
\ee
\be\label{19}
Q_{\phi} = \int_{-\infty}^{+\infty} dx~\phi^{\dag} \phi\,.
\ee

\subsection{Approximate Solution}

We will obtain our approximate analytic solution by  assuming that each of 
the two terms in Eq. (\ref{15}) is identically zero.  Then we obtain

\bq\label{20}
\frac{d \theta}{d x}  = -   ( \omega_1 - m_1  \cos 2 \theta )\,, ~~
\frac{d \eta}{d x}  = -   ( \omega_2 - m_2  \cos 2 \eta )\,, 
\eq
whose solutions are the same as when $g_3=0$, namely
\bq \label{21}
\theta(x) = \tan^{-1} (\alpha_1 \tanh \beta_1 (x),  ~~\eta(x) = \tan^{-1} (\alpha_2 \tanh \beta_2 (x),
\eq
where 
\bq\label{22}
\alpha_i = \sqrt{\frac{m_i- \omega_i}{m_i+\omega_1}} , ~~
\beta_i=   \sqrt{m_i^2-\omega_i^2}. 
\eq

Using  Eqs. (\ref{8}), (\ref{20}) and (\ref{21}) we can solve for $y_1$ 
and $y_2$ and obtain
\ba\label{23}
&&y_1= 2 \sec 2\theta   \frac{[-g_2^2 \omega_1 \sec 2\theta  +g_2^2
m_1+g_3^2 \omega_2 \sec 2 \eta -g_3^2 m_2]} {g_1^2 g_2^2-g_3^4} \nonumber \\
&&=\frac{1}{(1-g_{32}^2 g_{31}^2) }  \left[ y_{10} - g_{31}^2 y_{20} 
\frac{\sec 2 \theta}{\sec 2 \eta} \right] , 
\ea
\ba\label{23a}
&&y_2 =  2 \sec 2 \eta \frac{ [-g_1^2 \omega_2 \sec 2 \eta +g_1^2
m_2+g_3^2 \omega_1 \sec 2 \theta -g_3^2 m_1]} {g_1^2 g_2^2-g_3^4} \nonumber \\
&&=\frac{1}{(1-g_{32}^2 g_{31}^2) }  \left[ y_{20} - g_{32}^2 y_{10} 
\frac{\sec 2 \eta}{\sec 2 \theta} \right]   , 
\ea
where  $y_{i0}$ is the value of $y_i$ when $g_3=0$, and $g_{3i} =g_{3}/g_{i}$ 
for $i=1,2$. Letting   $\theta_i = (\theta, \eta)$ we have that  the values 
of various trigonometric functions of $\theta_i$, valid when $g_3=0$ are 
given by:

\bq\label{24}
\tan \theta_i  = \frac{\sqrt{m_i-\omega_i } \tanh 
\left(x \sqrt{m_i^2-\omega_i ^2}\right)}{\sqrt{m_i+\omega_i}} , 
\eq
\bq\label{25}
\sec  2 \theta_i = \frac{\omega_i 
+ m_i \cosh 2 \beta_i x} { m_i + \omega_i \cosh 2 \beta_i x} , 
\eq
\bq\label{26}
 \sin 2 \theta _i 
= \frac{\beta_i \sinh  2  \beta_i x }{m_i \cosh 2  \beta_i x  +\omega_i} . 
\eq
   
\begin{figure}
  \centering
  \includegraphics[width=7cm]{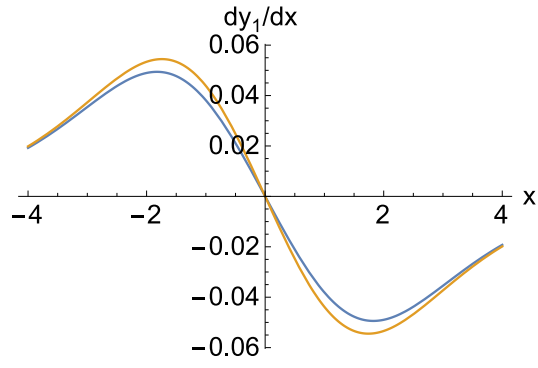}
  \caption{lhs (blue- curve)  vs. rhs of Eq. (\ref{8}) for$dy_1/dx$ 
when  $g_3^2 = 1/10$. }
  \label{dydxtest3}
\end{figure}

\begin{figure}
  \centering
  \includegraphics[width=7cm]{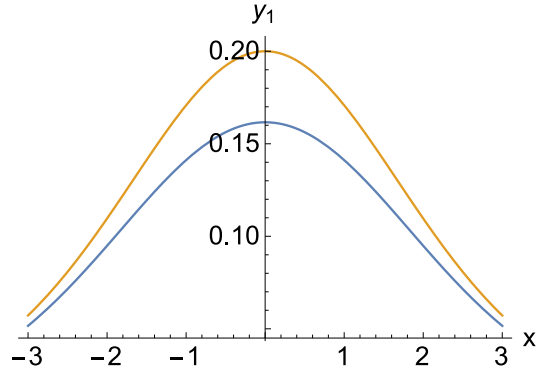}
  \caption{ $y_1$ (upper curve) and $y_{10}$ when $g_3^2=1/10$.}
  \label{y1s}
\end{figure}

\begin{figure}
  \centering
  \includegraphics[width=7cm]{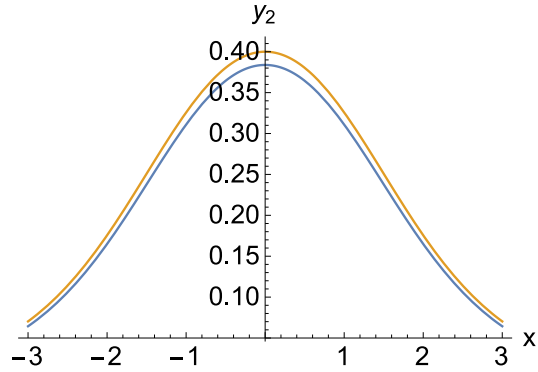}
  \caption{ $y_2$ (upper curve) and $y_{20}$  when $g_3^2=1/10$.}
  \label{y2s}
\end{figure}

We see from Eqs. (\ref{23}) and (\ref{23a}) that the solution for $y_i$  
after rescaling 
depends on the two dimensionless coupling constants $g_{3i}^2$. Our solutions 
for $y_i$ were found using the differential equation for $\theta_i$. So to 
see the accuracy of this solution we need to check how well the 
Eqs. (\ref{8}) are satisfied. We will find that this depends on whether 
one of the solutions $y_{i0}$ is double humped, since then its
 derivative near $x=0$ will be opposite that of the single humped one and 
then the left hand side can have behavior different from the right hand 
side in various scenarios near $x=0$. First, let us look at the case when 
both $y_{i0}$ are single humped. We can always after rescaling take 
$g_1=1$. For simplicity we will also choose $g_2=1$, and are then left with $g_{32} = g_{31} = g_3$ 
which for for illustrative purposes we will usually choose $= \sqrt {1/10}$.  For the choice 
$m_1=1$, $m_2=9/10$, $\omega_1= 9/10$, $\omega_2=7/10$, $g_3^2 = 1/10 $  
the left hand side of Eq. (\ref{8}) is shown in blue and the right hand side in yellow 
in Fig. \ref{dydxtest3}. 
Here if we take the ratio $(lhs-rhs)/(lhs+rhs)$ we would find that this 
was always less than $2 \%$ over the entire $x$ range. 
For this choice of parameters $y_1$ is much more modified by the interaction 
than $y_2$.  This is shown in Figs. \ref{y1s} and  \ref{y2s}.

\begin{figure}
  \centering
  \includegraphics[width=7cm]{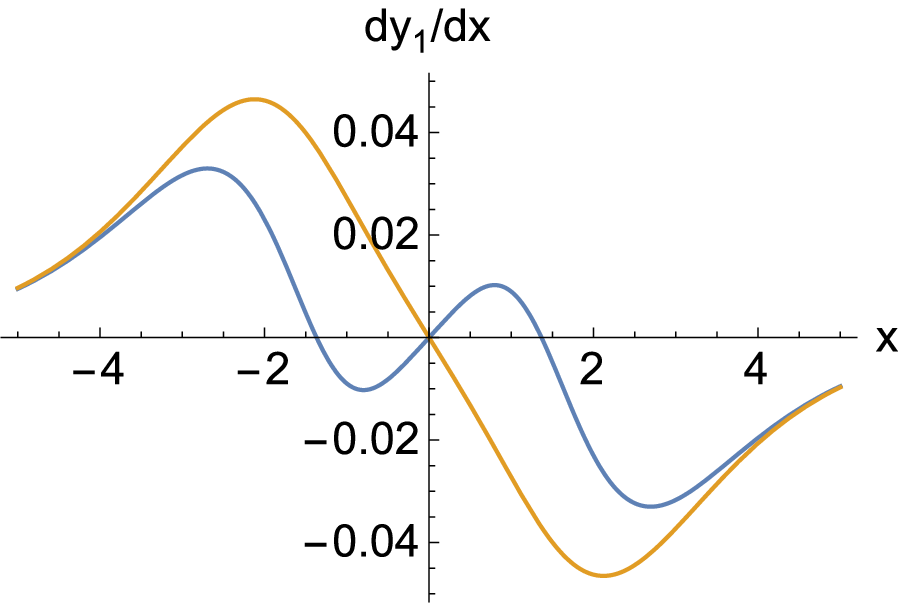}
  \caption{lhs (blue)  vs. rhs of Eq. (\ref{8}) for $dy_1/dx$ when $g_3^2=1/10$.}
  \label{bad}
\end{figure}

The problematic case is when one of the $y_{i0}$ is double humped. In that case 
the derivatives of $y_i$ determined from our approximation can be positive, 
whereas the right hand side can still be negative in our approximation for a 
range of small $x$ and $g_3^2 = 1/10$.  An example of this is given for the values:
$m_1=1$, $m_2=9/10$, $\omega_1= 9/10$, $\omega_2=3/10$, $g_3^2 = 1/10$ in Fig.  \ref{bad}

\begin{figure}
  \centering
  \includegraphics[width=7cm]{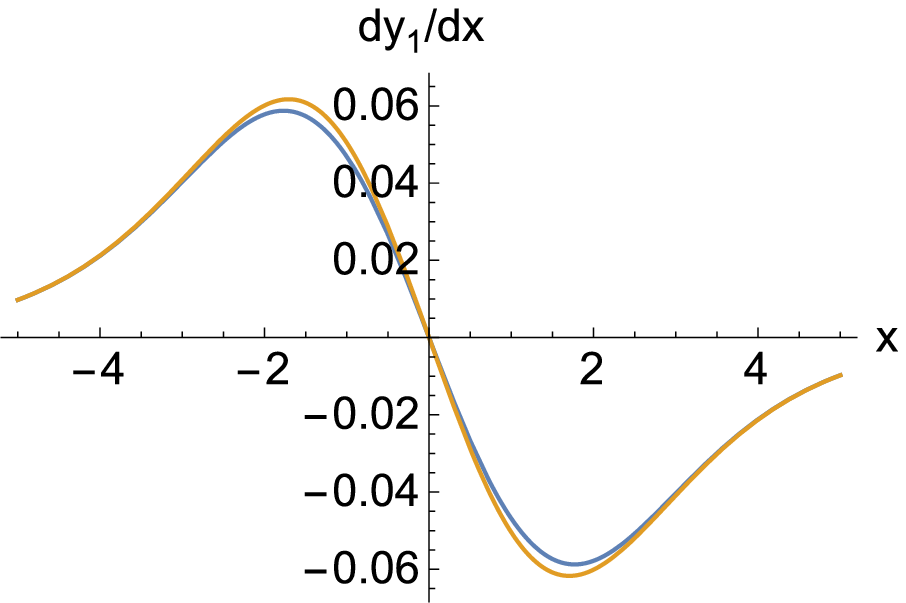}
  \caption{lhs vs. rhs of Eq. (\ref{8}) for $dy_1/dx$ when $g_3^2=1/100$.}
  \label{good}
\end{figure}

\begin{figure}
  \centering
  \includegraphics[width=7cm]{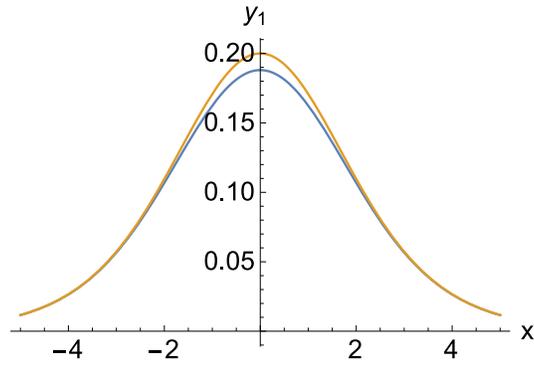}
  \caption{ $y_1$ when $g_3=0$  (upper curve) vs.   $y_1$ when $g_3^2=1/100$.}
  \label{y1good}
\end{figure}

\begin{figure}
  \centering
  \includegraphics[width=7cm]{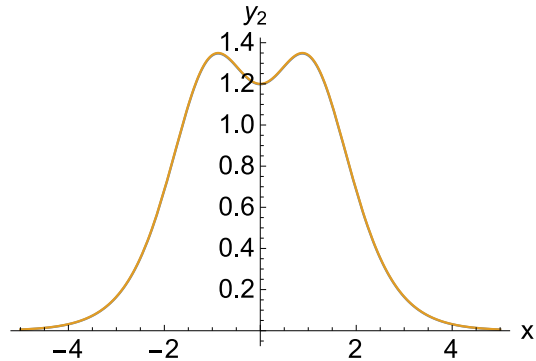}
  \caption{ $y_2$ when $g_3=0$   vs.    $y_2$ when $g_3^2=1/100$ (no visible difference at this scale.)}
  \label{y2good}
\end{figure}

In this case if we reduce $g_3^2$ to be $1/100$, then we again get good 
agreement between the left and right hand sides of the equation for $dy_1/dx$.
This is shown in  Fig.  \ref{good}.  When $g_3^2 = 1/100$,  $y_1$ is 
slightly modified from its value when $g_3=0$.  However, the double humped  
$y_2$ solution is barely modified by the interaction as seen in 
Figs. \ref{y1good} and \ref{y2good}.


\section{ Vector-Vector Interactions}

The coupled nonlinear Dirac 
equations (NLDEs) with vector-vector interactions  are given by 
\bq\label{3.1}
 (i \gamma^{\mu} \partial_{\mu} - m_1) \psi +g_1^2 \gamma^{\mu} \psi 
(\bpsi \gamma_{\mu}  \psi) 
+ g_3^2 \gamma^{\mu} \psi (\bphi \gamma_{\mu}  \phi) = 0\,, 
\eq
\bq\label{3.2}
(i \gamma^{\mu} \partial_{\mu} - m_2) \phi +g_2^2 \gamma^{\mu} \phi 
(\bphi \gamma_{\mu}  \phi) + g_3^2 \gamma^{\mu} \phi (\bpsi \gamma_{\mu}
 \psi) =0\,.
\eq
Again by scaling $\psi \rightarrow  \psi/g_1$,   $\phi \rightarrow  \phi/g_2$, we have only two
independent dimensionless coupling  constants  $g_3^2/ g_2^2  $ and  
$g_3^2/ g_1^2$.  Equations (\ref{3.1})  and (\ref{3.2}) can be derived from the 
Lagrangian
\ba\label{3.3}
\mathcal{L} && =  \bpsi (i \gamma^{\mu} \partial_{\mu} - m_1) \psi + \frac{g_1^2 }{2} (\bpsi \gamma_{\mu}  \psi)(\bpsi \gamma^{\mu} \psi)   \nonumber \\
&& 
+ \bphi (i \gamma^{\mu} \partial_{\mu} - m_2) \phi +\frac{g_2^2 }{2} 
(\bphi \gamma_{\mu} \phi)(\bphi \gamma^{\mu} \phi) + g_3^2 (\bpsi \gamma_{\mu} \psi)(\bphi \gamma^{\mu}  \phi ) \nonumber \\
&&= \bpsi (i \gamma^{\mu} \partial_{\mu}- m_1) \psi + \bphi (i \gamma^{\mu} \partial_{\mu} - m_2) \phi + \mathcal{L} _{int} . 
\ea
We notice that as in the scalar-scalar case, the Lagrangian in this case is 
also symmetric under the interchange $\psi \rightarrow  \phi$, 
$m_1 \rightarrow m_2$ and $g_1 \rightarrow g_2$. Again 
using the representation as given by Eq. (\ref{6}), we have the 
equations for the components of the two coupled NLDEs which can be written as

%
%

\ba\label{3.4}
&&\partial_x A + (m_1 +\omega_1) B +g_1^2 (A^2+B^2)  B +g_3^2 (C^2+D^2)  B  =0, \nonumber \\ 
&&\partial_x B + (m_1-\omega_1) A -g_1^2 (A^2+B^2)  A - g_3^2 (C^2+D^2)  A =0, \nonumber \\
&&\partial_x C + (m_2 +\omega_2) D + g_2^2 (C^2+D^2)  D +g_3^2 (A^2+B^2)  D  =0, \nonumber \\ 
&&\partial_x D + (m_2 +\omega_2) C  -g_2^2 (C^2+D^2)  C  -g_3^2 (A^2+B^2)  C  =0 .
\ea
These are symmetric under the interchange $ \{ A, B\} \rightarrow  \{ C, D\}$, $m_1 \rightarrow m_2$ and $g_1 \rightarrow g_2$.
These four  equations  can also be written if we let $y_i=R_i^2(x)$ as:
\ba\label{3.5} 
\frac{dy_1}{dx} && = -2 y_1 m_1\sin 2 \theta , 
 \nonumber \\
\frac{dy_2}{dx} && = -2 y_2 m_2 \sin 2 \eta  , 
\ea
and
\ba\label{3.6}
\frac{d \theta}{dx} &&= \omega_1+ g_1^2 y_1 +  g_3^2 y_2   -m_1  \cos   2 \theta  , 
 \nonumber  \\
\frac{d \eta}{dx} &&=\omega_2  +g_2^2 y_2 +  g_3^2 y_1 -m_2  \cos   2 \eta  . 
\ea
This reduces, when $g_3=0$  to the Eqs. (14) in Chang et al.  \cite{Cha}.
%

\subsection{Conservation Laws}

We again have energy-momentum conservation governed by Eqs. 
(\ref{10}) and (\ref{11}) but where $\mathcal{L}$ is now given by 
Eq. (\ref{3.3}).
From total momentum conservation,  we find, just like for the scalar case, 
that for a solution that vanishes at $\pm \infty$, $T_{10}$ and $T_{11}$ 
are again given by Eqs. (\ref{12}) and (\ref{13}) respectively, but where
$\mathcal{L}_{int}$ is as given by Eq. (\ref{3.3}).  
Multiplying Eq. (\ref{3.1}) on the left by $\bpsi$ and Eq. (\ref{3.2})
 on the left by $\bphi$  and adding those two equations and then using  
Eq. (\ref{13}) to eliminate the interaction terms of $\mathcal{L}_{int}$, 
we as in the scalar case, again obtain Eq. (\ref{14}). On using the ansatz
(\ref{6}) we then again obtain  
\bq \label{3.7}
R_1^2 (\frac{d\theta}{dx} + \omega_1 - m_1 \cos 2 \theta) 
+ R_2^2 (\frac{d\eta}{dx} + \omega_2 - m_2 \cos 2 \eta) =0\,.
\eq
As in the scalar case, in the vector case too the energy and the charges
$Q_{\psi}$ and $Q_{\phi}$ are conserved and are again given by
Eqs. (\ref{17}) to (\ref{19}), respectively.

\subsection{Approximate Solution}

We will obtain our approximate analytic solution by  assuming that each of 
the two terms in Eq. (\ref{3.7}) is identically zero.  Then we obtain
\bq  \label{3.8}
\frac{d \theta}{d x}  =   ( m_1  \cos 2 \theta - \omega_1), ~~
\frac{d \eta}{d x}  =  (m_2  \cos 2 \eta - \omega_2) , 
\eq
whose solutions are given by Eq. (\ref{20}).
%
We can again solve for $y_1$ and $y_2$ and obtain
\ba \label{3.9}
y_1 &&= \frac{2 \left({g_2}^2 {m_1} \cos (2 \theta )-{g_2}^2 \omega_1-{g_3}^2
   m_2 \cos (2 \eta )+{g_3}^2 \omega_2 \right)}{{g_1}^2 , 
   {g_2}^2-{g_3}^4 } \nonumber \\
 y_2 &&=  \frac{2 \left({g_1}^2 {m_2} \cos (2 \eta )-{g_1}^2 \omega_2-{g_3}^2
   m_1 \cos (2 \theta )+{g_3}^2 \omega_1 \right)}{{g_1}^2 . 
   {g_2}^2-{g_3}^4 }
 \ea
 Since in the absence of interactions $(g_3=0)$ we have
 \bq\label{3.10}
 y_{10} =\frac{2}{{g_1}^2} \left[ m_1 \cos (2 \theta)- \omega_1 \right] , ~~~y_{20} =\frac{2}{{g_2}^2} \left[ m_2 \cos (2 \eta) - \omega_2  \right] , 
 \eq
 we can rewrite Eq. (\ref{3.9}) as
 
 \ba \label{3.11}
y_1 &&=  \frac{1}{1-g_{31}^2 g_{32}^2}  
\left[y_{10}- g_{31}^2 y_{20} \right]   ,    \nonumber \\
 y_2 &&=  \frac{1}{1-g_{31}^2 g_{32}^2}  
\left[y_{20}- g_{32}^2 y_{10} \right]    . 
    \ea
So we see that we need both $g_{31}^2 \ll1$ and  $g_{32}^2 \ll 1$ for this 
approximation to make sense. Now let us see to what extent we violate 
Eq. \ref{3.5}. We have, letting $\theta_i = (\theta, \eta)$, the approximate 
expression for $\theta_i$, valid when $g_3=0$, given by Eq. (\ref{20}).

Now unlike the scalar-scalar case, the solutions $y_{i0}$ are single humped and so typical values of the parameters give generic results.

\begin{figure}
  \centering
  \includegraphics[width=7cm]{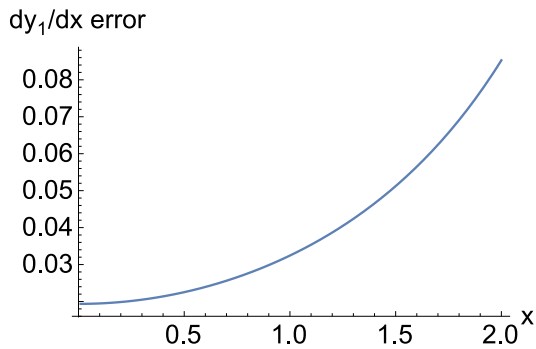}
  \caption{(lhs-rhs)/(lhs+rhs)  of Eq. (\ref{3.5}0 for $dy_1/dx$ when $g_3^2=1/10$.}
  \label{testvv}
\end{figure}

\begin{figure}
  \centering
  \includegraphics[width=7cm]{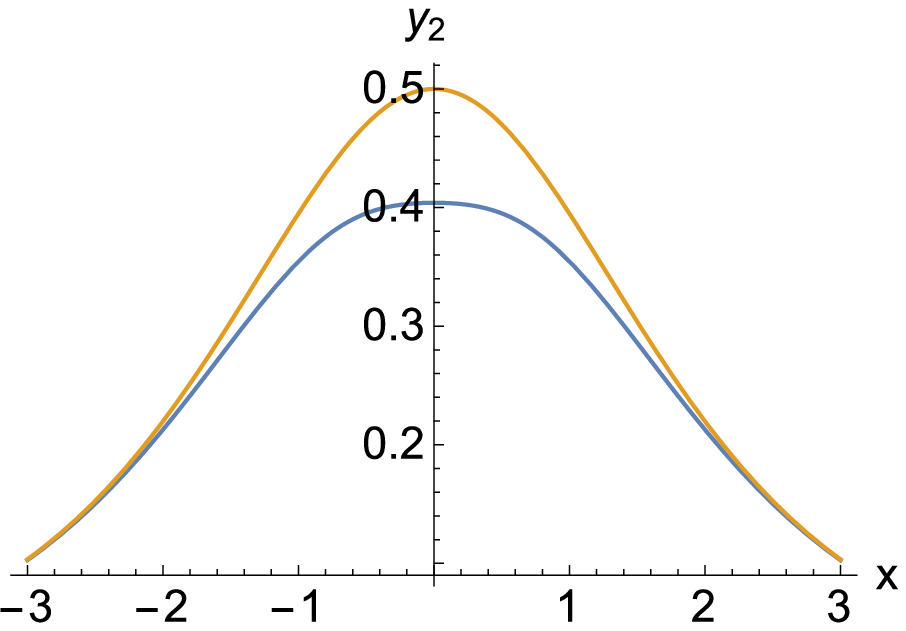}
  \caption{ $y_2$ (upper curve) and $y_{20}$  when $g_3^2=1/10$.}
  \label{y2vv}
\end{figure}

\begin{figure}
  \centering
  \includegraphics[width=7cm]{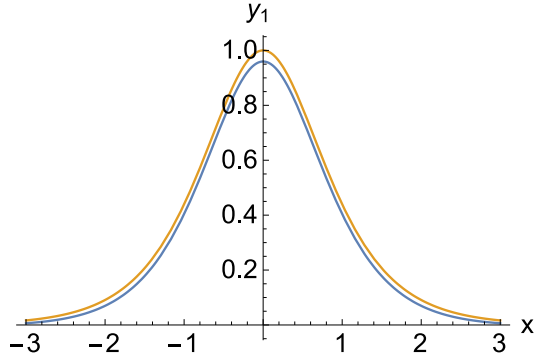}
  \caption{ $y_1$ (upper curve) and $y_{10}$ when $g_3^2=1/10$.}
  \label{y1vv}
\end{figure}

Setting $g_1=g_2=1$ and $g_3^2 = 1/10$  and $m_1=1$, $m_2 =1/2$, 
$\omega_1  =1/2$, $\omega_2=  1/4$ we find that the relative error on comparing 
the lhs and rhs of  Eq. (\ref{3.5}), i.e $(lhs-rhs)/(lhs+rhs)$, is less than 
  $3 \%$, (see Fig. \ref{testvv} ) for $ |x| <1$,  At the same time, $y_2$ 
is changed quite a bit from its uncoupled value when we choose these values 
of the parameters as seen in Fig. \ref{y2vv}. The effect is not as dramatic 
for $y_1$ for  these values as seen in Fig. \ref{y1vv}.


\section{Nonrelativistic Limit}

In our previous paper \cite{coo}, we had started with NLD equations and using
Moore's decoupling method \cite{moo} we had obtained the nonrelativistic limit
of our NLD equations in both the scalar and the vector coupling cases.
In this section we essentially follow the same decoupling method to obtain
the nonrelativistic limit of coupled NLD equations in both scalar and
vector coupling cases.
 
Let us start from the coupled Eqs. (\ref{1}) and (\ref{2}) or 
(\ref{3.1}) and (\ref{3.2}). They can be reexpressed in the form
\bq\label{4.1}
i \sigma_3 \partial_{t} \psi + \sigma_x \partial_{x} \psi  - m_1 \psi 
- V_{I}^{1} \psi = 0\,,
\eq
\bq\label{4.2}
i \sigma_3 \partial_{t} \phi + \sigma_x \partial_{x} \phi - m_2 \phi 
- V_{I}^{2} \phi = 0\,,
\eq
where $V_{I}^{1} = -\frac{\partial L_{I}}{\partial \psi}$ while
$V_{I}^{2} = -\frac{\partial L_{I}}{\partial \phi}$ 
and $L_{I}$ is as given by Eq. (\ref{4}) or (\ref{3.3}). 
On using 
\be\label{4.4}
\psi = \left(  \begin{array} {cc}
      u_0 \\
      w_0 \\ 
   \end{array} \right) e^{-i \omega_1 t}\,,
   \eq
\bq\label{4.5}
\phi = \left(  \begin{array} {cc}
      v_0 \\
      z_0 \\ 
   \end{array} \right) e^{-i \omega_2 t}\,,
   \eq
and the Moore's decoupling method as well as essentially following the steps
given in our previous paper \cite{coo}, we find that in both the 
scalar-scalar  and the vector-vector cases we get the coupled 
NLS equations
\bq\label{4.6}
-(u_{0})_{xx} +(m_{1}^{2} - \omega_{1}^{2}) u_0 -(m_1+\omega_1)
[g_{1}^{2} |u_0|^2 + g_{3}^{2} |v_0|^2] u_0 = 0\,,
\eq
\bq\label{4.7}
-(v_{0})_{xx} +(m_{2}^{2} - \omega_{2}^{2}) v_0 -(m_2+\omega_2)
[g_{2}^{2} |v_0|^2 + g_{3}^{2} |u_0|^2] v_0 = 0\,,
\eq
under the assumption that $(m_1 - \omega_1)/2m_1 \ll1$ and 
$(m_2 - \omega_2)/2m_2 \ll 1$. 

Let us now look for exact solutions of the coupled Eqs. (\ref{4.6}) and
(\ref{4.7}) under the assumption that both $u_0$ and $v_0$ vanish in the 
limit $x \rightarrow \pm \infty$. It turns out that there are two such
solutions and we discuss these one by one. 

\subsection{Solution I}

It is easy to check that 
\be\label{4.8}
u_0 = A \sech \beta x\,,~~v_0 = B \sech \beta x\,,
\eq
is an exact solution to the coupled Eqs. (\ref{4.6}) and (\ref{4.7}) provided
\bq\label{4.9}
m_{1}^{2} - \omega_{1}^{2} = m_{2}^{2} - \omega_{2}^{2} = \beta^2\,,
\eq
\bq\label{4.10}
g_{1}^{2} A^2 + g_{3}^{2} B^2 = 2(m_1 - \omega_1)\,,
\eq
\bq\label{4.11}
g_{2}^{2} B^2 + g_{3}^{2} A^2 = 2(m_2 - \omega_2)\,.
\eq
On solving Eqs. (\ref{4.10}) and (\ref{4.11}) we find that
\bq\label{4.12}
A^2 = \frac{2[(m_1-\omega_1)g_{2}^{2} - (m_2-\omega_2)g_{3}^{2}]}
{g_{1}^{2} g_{2}^{2} - g_{3}^{4}}\,,~~
B^2 = \frac{2[(m_2-\omega_2)g_{1}^{2} - (m_1-\omega_1)g_{3}^{2}]}
{g_{1}^{2} g_{2}^{2} - g_{3}^{4}}\,,
\eq
provided $g_{1}^{2} g_{2}^{2} \ne g_{3}^{4}$. In case 
$g_{1} g_{2} = g_{3}^{2}$ then $A$ and $B$ remain undetermined and we
only have the constraint
\bq\label{4.13}
g_1 A^2 + g_2 B^2 = \frac{2(m_1-\omega_1)}{g_1} 
= \frac{2(m_2-\omega_2)}{g_2}\,.
\eq 

\subsection{Solution II}

Another solution to the coupled Eqs. (\ref{4.6}) and (\ref{4.7})  satisfying
the boundary condition 
$u_0, v_0 \rightarrow 0$ as $x \rightarrow \pm \infty$ is
\be\label{4.14}
u_0 = A \sech^2 \beta x\,,~~v_0 = B \sech \beta x \tanh \beta x\,,
\eq
provided
\bq\label{4.15}
m_{1}^{2} - \omega_{1}^{2} = 4 \beta^2\,,~~m_{2}^{2} - \omega_{2}^{2} 
= \beta^2\,, ~~g_1 g_2 = g_{3}^{2}\,,~~g_1 A^2 = g_2 B^2\,,
\eq
\bq\label{4.16}
g_{1} (m_1+\omega_1) = g_{2} (m_2 + \omega_2)\,,
~~(m_1+\omega_1)g_{1}^{2} A^2 = 6 \beta^2\,.
\ee

It is thus worth noting that while the first solution is valid for any values
of $g_1, g_2, g_3$, the second solution is only valid when 
$g_1 g_2 = g_{3}^{2}$.

\section{Conclusions}
In this paper we have introduced and initiated discussion about coupled NLD 
equations with both scalar-scalar and vector-vector interactions. In particular, 
we have given the first (approximate) analytic solitary wave solution to two 
coupled NLDEs for both scalar-scalar interactions and vector-vector interactions. 
These solutions are relevant in nonlinear optics \cite{bar} as well as for light solitons 
in waveguide arrays \cite{lon, dre, tra} among other applications in BECs and 
cosmology. Further,we have shown using the Moore's decoupling method that in 
the nonrelativisticlimit, NLDEs with both scalar-scalar and vector-vector interactions 
reduce tothe same coupled nonlinear Schr\"odinger equation (NLSE). We have 
obtained two exact pulse solutions to these coupled NLSE.  
Using the results found in \cite{coo}, one can extend these solutions to the 
case where the scalar-scalar as well as vector-vector interactions are taken 
to an arbitrary (nonlinearity) power $\kappa$.  We hope to address this issue 
as well as the question of stability of the solutions found here in the near future.  

\begin{acknowledgments}

This work was performed in part under the auspices of the U.S.  
Department of Energy.  F.C.  would like to thank the Santa Fe Institute  and 
the Center for Nonlinear Studies, Los Alamos National Laboratory, for its 
hospitality. A.K. is grateful to Indian National Science Academy (INSA) for
awarding him INSA Senior Scientist position at Savitribai Phule Pune
University, Pune, India.

\end{acknowledgments}

%
%
%
%
%
%
%
%
%
%
%
%

\end{document}